\documentclass[aps,pra,twocolumn,superscriptaddress,nolongbibliography]{revtex4-2}
\usepackage[colorlinks=true, citecolor=blue,allcolors=blue]{hyperref}
\usepackage{graphicx}
\usepackage{dcolumn}
\usepackage{bm}
\usepackage{amsmath}

\begin{document}
\title{Magnetic Dichroism in Few-Photon Ionization of Polarized Atoms} 

\author{ B.P. Acharya}
\affiliation{Physics Department and LAMOR, Missouri University of Science \& Technology, Rolla, MO 65409, USA}

\author{M. Dodson}
\affiliation{Department of Physics, Kennesaw State University, Kennesaw, Georgia 30144, USA}

\author{S.\ Dubey}
\affiliation{Physics Department and LAMOR, Missouri University of Science \& Technology, Rolla, MO 65409, USA}

\author{K.L.\ Romans}
\affiliation{Physics Department and LAMOR, Missouri University of Science \& Technology, Rolla, MO 65409, USA}

\author{A.H.N.C.\ De~Silva}
\affiliation{Physics Department and LAMOR, Missouri University of Science \& Technology, Rolla, MO 65409, USA}

\author{ K.\ Foster}
\affiliation{Physics Department and LAMOR, Missouri University of Science \& Technology, Rolla, MO 65409, USA}

\author{O.\ Russ}
\affiliation{Physics Department and LAMOR, Missouri University of Science \& Technology, Rolla, MO 65409, USA}

\author{K.\ Bartschat}
\affiliation{Department of Physics  and Astronomy, Drake University, Des Moines, Iowa 50311, USA}

\author{N.\ Douguet}
\affiliation{Department of Physics, Kennesaw State University, Kennesaw, Georgia 30144, USA}

\author{D.\ Fischer}
\affiliation{Physics Department and LAMOR, Missouri University of Science \& Technology, Rolla, MO 65409, USA}

\date{\today}
             
\begin{abstract}
We consider few-photon ionization of atomic lithium by linearly polarized femtosecond laser pulses, and demonstrate 
that asymmetries of the electron angular distribution can occur for initially polarized (2$p$, $m$=+1) target atoms. 
The dependence of the photoelectron emission angle relative to the electric field direction is investigated
at different laser intensities and wavelengths. The experimental spectra show excellent
agreement with numerical solutions of the time-dependent 
Schr{\"o}dinger equation. In the perturbative picture, the angular shift is traced back to 
interferences between partial waves with mean magnetic quantum number $\left< m\right>\ne 0$.
This observation allows us to obtain quantum mechanical information on the final electronic 
state. 
\end{abstract}



\maketitle

\section{Introduction}

Atomic ionization in optical fields proceeds predominantly through the electric dipole interaction of the 
initially bound atomic system with the external field. Consequently, photoelectron angular distributions (PADs) 
are generally governed by the direction and symmetries of the electric field. In the simplest case of an 
unpolarized target, which is ionized by linearly or circularly polarized light, the symmetries of the electronic 
final state are (in the electric dipole approximation) identical to the symmetries of the ionizing field given by 
its Stokes parameters. However, there are more complex situations where these symmetries are lifted and the electron 
emission does not geometrically align with the dominant electric field direction.

Examples, which have been debated extensively in the past decade, are ``attoclock" 
experiments \cite{Eckle2008, Pfeiffer2011,Landsman2014,Camus2017,Sainadh2019}, 
where adiabatic tunnel ionization of atoms in elliptically 
polarized few-cycle pulses is investigated. In these measurements, the electron angular distributions, 
in the plane perpendicular to the laser propagation direction (i.e., in the \emph{azimuthal} plane),
feature a shift from the direction of the potential vector at the instant of strongest electric field. This shift in the azimuthal 
angle $\varphi$ might (partially) be attributed to a time delay of the ionization while the electron 
tunnels through the barrier formed by the potential of the atomic core and the adiabatically 
changing electric field of the laser. Although this interpretation is still somewhat controversial and the 
debate about the tunneling time remains open (for recent reviews, 
see \cite{Kheifets2020, SatyaSainadh2020}), joint experimental and theoretical efforts 
resulted in a much better understanding of the tunneling dynamics and an improved modeling 
of the complex strong-field--atom interaction. 

Already two decades before the first attoclock experiments, a related phenomenon was 
observed in the multi-photon ionization regime -- the so-called ``elliptic dichroism''  \cite{Bashkansky1988}. 
Here, again, the major and minor axes of the polarization ellipse do not represent lines of reflection symmetry 
in PADs measured in noble-gas ionization by elliptically polarized light. While the observed symmetry breaks 
are in contradiction to Keldysh-type theories \cite{Bashkansky1988, Keldysh1965, Faisal1973, Reiss1980}, 
they are qualitatively explained in terms of lowest-order perturbation theory (LOPT) \cite{Lambropoulos1988,Muller1988}. 
In this description, the asymmetry in the azimuthal electron emission angle $\varphi$ is a result of the interference 
of phase-shifted partial waves with different angular-momentum quantum numbers $\ell$ and $m$.

In the decades following the original discovery, elliptic dichroism attracted considerable interest and was observed, for instance, 
in above-threshold ionization of rare-gas targets \cite{Paulus1998,Paulus2000} as well as in few-photon 
ionization of alkali atoms \cite{Wang2000}. In contrast to the ionization by purely  linearly or circularly 
polarized light, analyzing ionization data for elliptic polarization enables to extract phases and 
amplitudes of the final partial waves, thereby allowing us to obtain the \emph{complete} quantum-mechanical 
information of the scattering process \cite{Dulieu1995,Wang2000b}. Recently, it was predicted that 
maximum elliptic dichroism can be achieved in two-photon ionization for an appropriate choice of 
radiation wavelength, thus making it a promising tool, e.g., to analyze the polarization state 
of free-electron laser radiation \cite{Hofbrucker2018}. It is worth noting that the ellipticity of 
the polarization is not a \emph{sine qua non} for angular asymmetries to occur. Similar asymmetric 
final states are expected, e.g., in multi-photon ionization by two combined laser beams of 
different colors; one with linear and the other one with circular polarization \cite{Taieb2000}.
 
In the present study, we demonstrate that left-right asymmetries can already be generated in atomic 
ionization by purely linearly polarized light if the target atoms are initially polarized. 
On the experimental side, it has been shown previously that optical traps are an ideal tool 
to provide excited and polarized atomic targets for ion-atom scattering \cite{Leredde2013,Hubele2013} 
or photoionization experiments \cite{Zhu2009,Thini2020,Silva2021}. Here we use an all-optical atom 
trap (AOT) \cite{Sharma2018} to prepare an excited lithium target in the polarized 2$p$ configuration with $m=+1$. 
The atoms are ionized by femto\-second laser pulses with a variable wavelength between 695\,nm and 800\,nm. 
We observe strong \emph{magnetic} dichroism, i.e., a dependence of the differential cross sections on the 
magnetic quantum number of the initial state \cite{Meyer2011}, which manifests itself as an angular shift of 
the main electron emission directions with respect to the laser polarization axis. The measured spectra are 
well reproduced by our calculations based on the numerical solution of the time-dependent Schr{\"o}dinger 
equation (TDSE), and strongly depend on both the intensity and wavelength of the laser pulse.   

The observed asymmetries are qualitatively explained in lowest-order perturbation theory (LOPT) and in
the dipole approximation, analogous to the discussions in \cite{Lambropoulos1988,Muller1988,Hofbrucker2018}. 
Despite its similarities to elliptic dichroism, the present scheme does not require non\-linear interactions 
with the laser field in order for asymmetries to appear \cite{Hofbrucker2018}, but they are, in principle, 
already present after the absorption of a single photon \cite{Thini2020}. Moreover, the present approach is expected, in the future, to
contribute to the ongoing discussion about tunneling times in attoclock experiments, because it might allow us to 
disentangle contributions to the angular shifts caused by the tunneling dynamics and by other effects such as, e.g., 
the long-range Coulomb interaction between the emitted electron and the photo-ion.

\section{Experiment}
Since the experimental setup has been described previously \cite{Thini2020,Silva2021, Silva2021c}, only a 
brief summary is given here. The experiment consists of three major components: (i) an optical trap providing 
state-prepared lithium target atoms, (ii) a tunable femtosecond laser source generating the ionizing 
external field, and (iii) a ``Reaction Microscope" measuring the momentum vectors of the ionization products.

The lithium target cloud is prepared in a near-resonant AOT \cite{Sharma2018}, 
where the atoms are cooled to temperatures in the milli-Kelvin range and confined to a small volume of 
about 1\,mm diameter. The cooling laser system consists of an external cavity diode laser with a tapered 
amplifier, whose frequency is stabilized near the $^{6}$Li $D2$-transition at about $\lambda = 671\,$nm. 
The radiation couples the $(2s)^2S_{1/2}$ to the $(2p)^2P_{3/2}$ state, and -- in steady state -- about 25\,\% of 
the target atoms populate the excited $P$ level, with about 93\,\% of them being in a single magnetic 
sub-state with $m= +1$ with respect to the direction of a weak magnetic field (the $z$-direction).

The femto\-second laser source is a commercially available system based on a Ti:Sa oscillator with two 
non-collinear optical parametric amplifier (NOPA) stages (e.g.~\cite{Harth2017}) providing maximum pulse 
energies of up to 15\,$\mu$J at a repetition rate of 200\,kHz. The system can be operated in a short-pulse 
(about 7\,fs FWHM of intensity)  broadband mode (ca.~660\,nm$-$1000\,nm). In the present experiment, however, we amplified only a 
rather narrow bandwidth ($\pm 15$\,nm) resulting in Fourier-limited pulse durations of about 35\,fs. The laser beam is 
guided into the vacuum chamber and focused into the lithium cloud with a minimum beam waist of about 50\,$\mu$m. 
The pulse duration and focal beam waist are used to estimate the pulses' peak intensities from the measured average power. 
In all the measurements performed in this study the peak intensity was between 1.0 and $4.0\times 10^{11}$\,W/cm$^2$.

A cold target recoil-ion momentum spectrometer (COLTRIMS) -- also referred to as 
``Reaction Microscope" \cite{Hubele2015,Fischer2019} -- is employed to measure the three-dimensional 
momentum vectors of both the electrons and recoil ions after the ionization process. A typical electron momentum resolution of 0.005 to 0.01\,a.u.\ is achieved \cite{Thini2020}. The differential cross section 
of the ionization of the Li(2$s$) ground state and of the Li(2$p$, $m=+1$) excited state are extracted employing a procedure that is described in more detail in \cite{Silva2021c}. In brief, the near-resonant cooling lasers are switched off periodically for short times. During these times, all target atoms are in the ground state and the data for Li(2$s$) ionization can be acquired. While the cooling lasers are switched on, a fraction of target atoms are in the excited state and the Li(2$p$, $m=+1$) ionization cross sections are obtained by subtracting the data for the cooling lasers being on and off using an appropriate scaling factor.

\section{Theory}
The experimental data are compared to {\it ab initio} calculations based on solving the  TDSE considering a single-active electron (SAE) in a He-like $1s^2$ ionic core. 
A static Hartree potential~\cite{Albright1993,Schuricke2011} is used and supplemented by phenomenological terms, 
which are discussed in \cite{Silva2021}. As shown earlier \cite{Silva2021c}, our model potential describes the 
atomic structure with an accuracy  better than 1~meV for the $n=2$ and $n=3$ states. Previous calculations 
using the same code yield excellent agreement with experimental data measured under similar conditions~\cite{Silva2021,Silva2021c}.

\section{Results and discussion}

\begin{figure}
\centering
\includegraphics[width=1\linewidth]{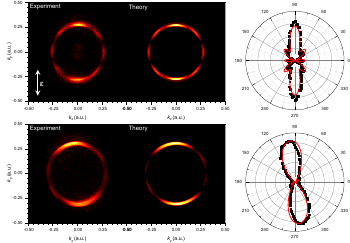}
\caption{Differential cross sections for few-photon ionization of lithium atoms initially in the 2$s$ (top) 
and 2$p$($m$=+1) (bottom) state in 35\,fs laser pulses at a center wavelength of 770\,nm and a peak intensity 
of 1.8$\times 10^{11}$\,W/cm$^2$. The initial 2$p$ state is polarized along the $z$ direction (perpendicular to 
the drawing plane), the laser field is polarized in the $y$ direction (i.e.\ vertically). Left and center 
columns show experimental and theoretical momentum distributions, respectively. The right column shows the 
distribution of the azimuthal angle. All spectra represent cuts in the $xy$ plane, i.e., $p_z=0$. In the calculations, 
we considered a lower intensity (by a factor of 1.8) than stated for the experiment, corresponding to an estimated 
mean intensity after averaging over the non-uniform spatial intensity distribution in the reaction volume (see text).\label{fig:770nm180mW}}
\end{figure}

In Fig.~\ref{fig:770nm180mW}, the momentum and angular distributions for the ionization of the initial 2$s$ 
and 2$p$ states are shown for a center wavelength of 770\,nm and a peak intensity of 1.8$\times10^{11}$\,W/cm$^2$. 
The laser field is polarized along the $y$ axis, and the orbital angular momentum of the excited $P$ state is 
polarized in the $z$ direction, perpendicular to the drawing plane in the figure. For all data presented in this study, the 
given laser field parameters resulted in Keldysh parameters well above 10, such that the ionization process 
can be described in the multi-photon picture. The initial 2$s$ state is ionized by the absorption of four photons, 
resulting in an asymptotic momentum $|\bm{p}|\approx0.28$\,a.u., which is reflected in a single-ring structure in the momentum distribution. 
The 2$p$-state ionization proceeds through the absorption of three photons corresponding to a slightly larger 
final state momentum of about $|\bm{p}|\approx0.31$\,a.u.. For the ground-state ionization, the angular differential cross 
section is symmetric with respect to the laser polarization axis (the $y$ axis in the graph) with its highest 
intensity in the direction of the laser electric field at $\varphi=90^\circ$ and 270$^\circ$. On the other hand, 
this symmetry is noticeably broken for ionization of  the $2p$ state, as the peaks in the angular distribution are shifted 
away from the electric field direction by $\Delta\varphi\approx10^\circ$. 

For a rigorous comparison of the measured spectra with the TDSE simulations, the non-uniform spatial intensity 
distribution of the laser field around the focal point should be taken into account. In the experiment, 
the location of a specific ionization event is not precisely known and, therefore, our experimental data are not 
measured for a well-defined intensity, but averaged over an intensity range. In previous studies, we had convolved 
the theoretical cross sections over a broad intensity range (e.g., \cite{Silva2021}), yielding nearly perfect 
agreement between measurements and calculations. While we expect 
that this procedure would reduce discrepancies, intensity-dependent features of the calculated spectra are more 
clearly visible without the averaging. Therefore, we omit this convolution in the present study and perform instead the calculation at a mean intensity by a 
factor 1.8 lower than the peak intensity applied in the experiment. Overall, the shape of the measured and 
calculated spectra are in excellent agreement (see Fig.~\ref{fig:770nm180mW}). 

\begin{figure}
\centering
\includegraphics[width=1\linewidth]{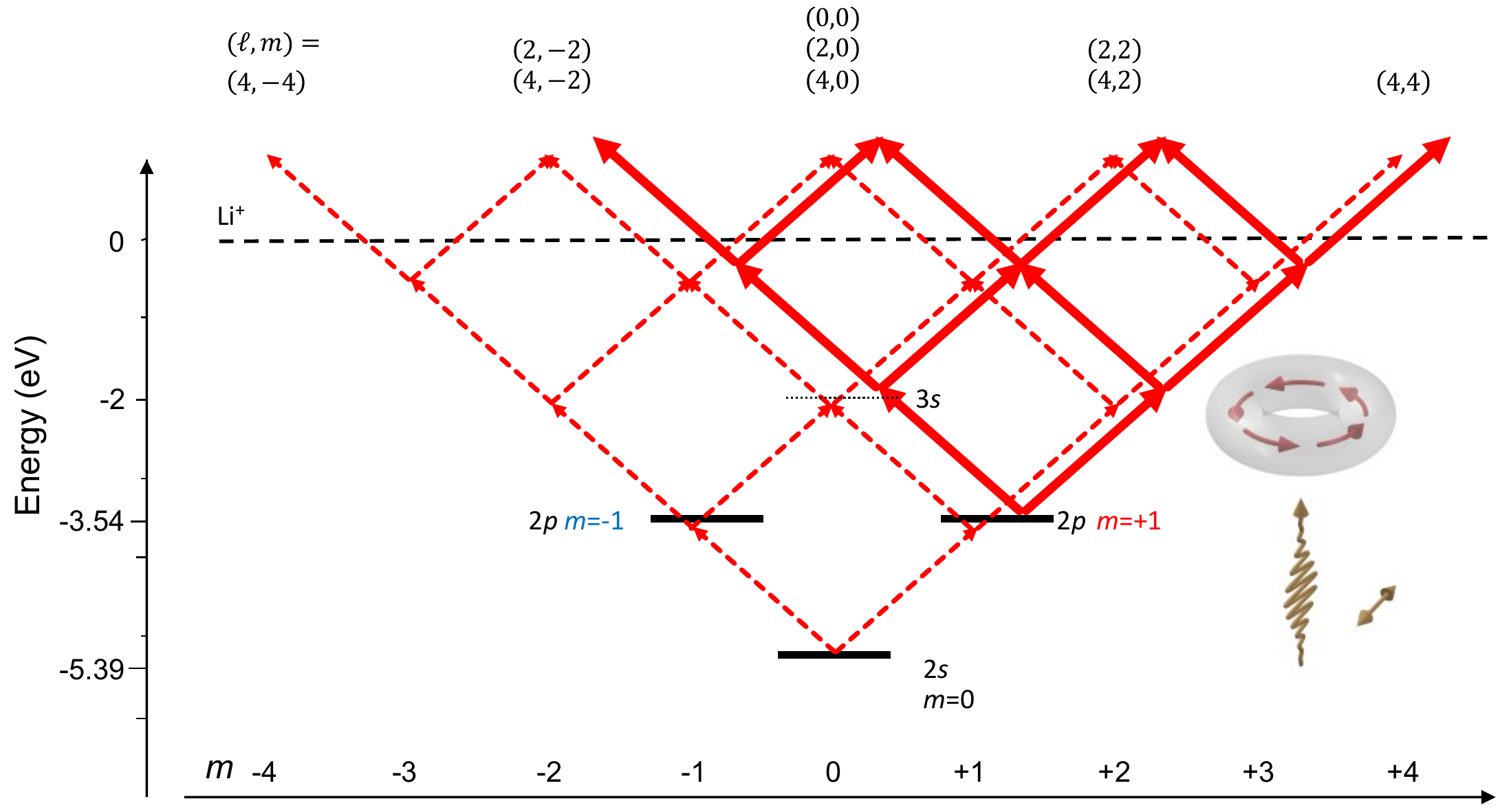}
\caption{Ionization scheme for three-photon ionization of the 2$p$ excited state (solid arrows) as 
well as for four-photon ionization of the 2$s$ ground state (dashed arrows) in a field with linear 
polarization oriented perpendicular to the atomic quantization direction. \label{fig:scheme}}
\end{figure}

The general features observed in the PADs can qualitatively be explained in the LOPT picture. For the following discussion, we choose a quantization direction along the $z$ axis, which coincides with the direction of the atomic polarization for the excited target initial state.
In the electric dipole approximation, the selection rules yield a change of the magnetic quantum number 
by $\Delta m=+1$ and $-1$ for each absorbed/emitted photon of right-handed and left-handed circular polarization in the $xy$ plane, respectively. Linearly 
polarized light along the $y$ direction corresponds to the coherent superposition of the two circular polarizations.
The resulting LOPT ionization pathways are depicted in Fig.~\ref{fig:scheme}. The angular part of the final electronic continuum 
state can be expressed in terms of a superposition of spherical harmonics $Y_{\ell m}(\vartheta,\varphi)$ of 
different dipole-allowed quantum numbers $\ell$ and $m$, which are -- for the given initial states, and depending on the number of 
absorbed and emitted photons -- either all even or all odd. In the presently considered case of 3-photon ionization 
of a 2$p~(m=+1)$ initial state, the allowed quantum numbers are $\ell=0$, 2, and 4 (corresponding 
to $s$, $d$, and $g$ waves) and $m=-2$, 0, 2, and~4.


The dependence of the final state wave function on the azimuthal angle $\varphi$ is generally given by (e.g.~\cite{Muller1988})
\begin{equation}
\Psi(\bm{k})=\sum_{\ell,m}a_{\ell m}(k,\vartheta) e^{im\varphi},
\end{equation}
with $a_{\ell m}$ relating to the complex amplitudes of the contributing partial waves, which generally depend on the absolute value of the photoelectron momentum $k$ and the polar emission angle $\vartheta$. 
For photoelectrons emitted in the $xy$ plane with an energy $E$, the above equation simplifies to
\begin{equation}
\Psi(k=\sqrt{2E},\vartheta=90^\circ,\varphi)=\sum_{m}c_{m} e^{im\varphi},
\label{eq:sumphi}
\end{equation}
with $c_m=\sum_\ell a_{\ell m}(k=\sqrt{2E},\vartheta=90^\circ)$ generally being complex. The photoelectron angular distribution is then obtained by the absolute square of this wave function:


\begin{widetext}
\begin{equation}
\left(\frac{\mathrm{d}\sigma}{\mathrm{d}\Omega}\right)_{\vartheta=90^\circ}=\left| \sum_{m}c_{m} e^{im\varphi}\right|^2
= \sum_{m,m'}c_{m}c^*_{m'}e^{i(m-m')\varphi}= \sum_{m}\left| c_{m}\right|^2+ 
\sum_{m<m'}2\left|c_m\right|\left|c_{m'}^*\right|\cos((m'-m)\varphi+\Delta_{m'm})\label{eq:dcs}.
\end{equation}
\end{widetext}

Any $\varphi$ dependence of the cross section stems from the interference of partial waves with different $m$
giving rise to interference terms that feature cosine functions oscillating with the emission angle $\varphi$ (see right-hand side of Eq.~(\ref{eq:dcs})). 
The angular shifts $\Delta_{m'm}$ correspond to the relative phase angles between the complex amplitudes $c_m$ and $c_{m'}$.
For the specific case of 2-photon ionization of the 2$p$ state shown in Fig.~\ref{fig:scheme}, 
the quantum number $m$ can take four values ($-2$, 0, 2, and 4). This results in a superposition of 
interference terms oscillating with $(m'-m)\varphi=2\varphi$, $4\varphi$, and $6\varphi$, which corresponds to an 
angular distribution with up to six local maxima in accordance with our data shown in Fig.~\ref{fig:770nm180mW} (bottom).

From Eq.~(\ref{eq:dcs}) it can be seen that the positions of the peaks in the angular distribution depend on the relative 
phase angles $\Delta_{m'm}$ of the interfering partial-wave amplitudes. The complex phases of these multi-photon amplitudes 
$\arg[c_m]$ are determined by several factors: First, there is a trivial dependence on the orientation of the laser field 
polarization. This  phase offset is (by our choice of the coordinate system) zero for a laser polarization in the $x$ direction and changes as the 
polarization is rotated in the $xy$ plane. Second, there are (asymptotic) phase shifts $\delta_\ell$ of the outgoing 
radial continuum wave functions, which are different for each $\ell$. Those include the well-known Coulomb phase shift 
$\arg\Gamma(\ell+1-i/k)$, but also non-Coulombic contributions due to the short-range part of the target potential. 
Third, bound and continuum intermediate states also have an effect on the multi-photon amplitude to a given partial wave, making their final phase generally path-dependent.

It is important to note that the photoelectron angular distribution can still be symmetric with respect to the photon polarization direction, even though individual interference terms generally do not feature this symmetry due to their rotation by the angle $\Delta_{m'm}$. A simple example is the 3-photon ionization of the 
lithium 2$s$ state shown in Fig.~\ref{fig:scheme} (dashed arrows), where the final state is composed of the orbital 
angular momenta $\ell=0$, 2, and 4 with $m$ ranging from $-4$ to $4$ \footnote{The 3-photon ionization of Li(2$s$) is more conveniently described for a quantization direction along the $y$ axis (parallel to the laser polarization direction). For this choice of the coordinate system, the dipole selection rules simplify to $\Delta m=0$, and the final state is a superposition of only three spherical harmonics with $\ell=0$, 2, and 4, all with $m=0$. The resulting photoelectron angular distribution depends only on the angle to the $y$-axis, which represents a symmetry axis.}. Here, each pair of partial waves with amplitudes $c_m$ and $c_{m'}$ possesses a counter pair $c_{-m}$ and $c_{-m'}$ such that their resulting interference terms 
are mirror images of one another with a reflection line given by the field's polarization. For this system, the superposition of all interference terms of Eq.~(\ref{eq:dcs}) results in a symmetric distribution. However, this symmetry between amplitudes of positive and negative $m$ is lifted, if there is a nonzero mean final-state polarization with $\left<m\right>\neq 0$.
This occurs, e.g., for a target in a polarized initial state with $m\neq 0$ (as in the present study), or if the target is ionized by 
elliptically polarized light. In this case, the angular symmetry with respect to the laser polarization direction is generally expected to be broken.

\begin{figure}
\centering
\includegraphics[width=1\linewidth]{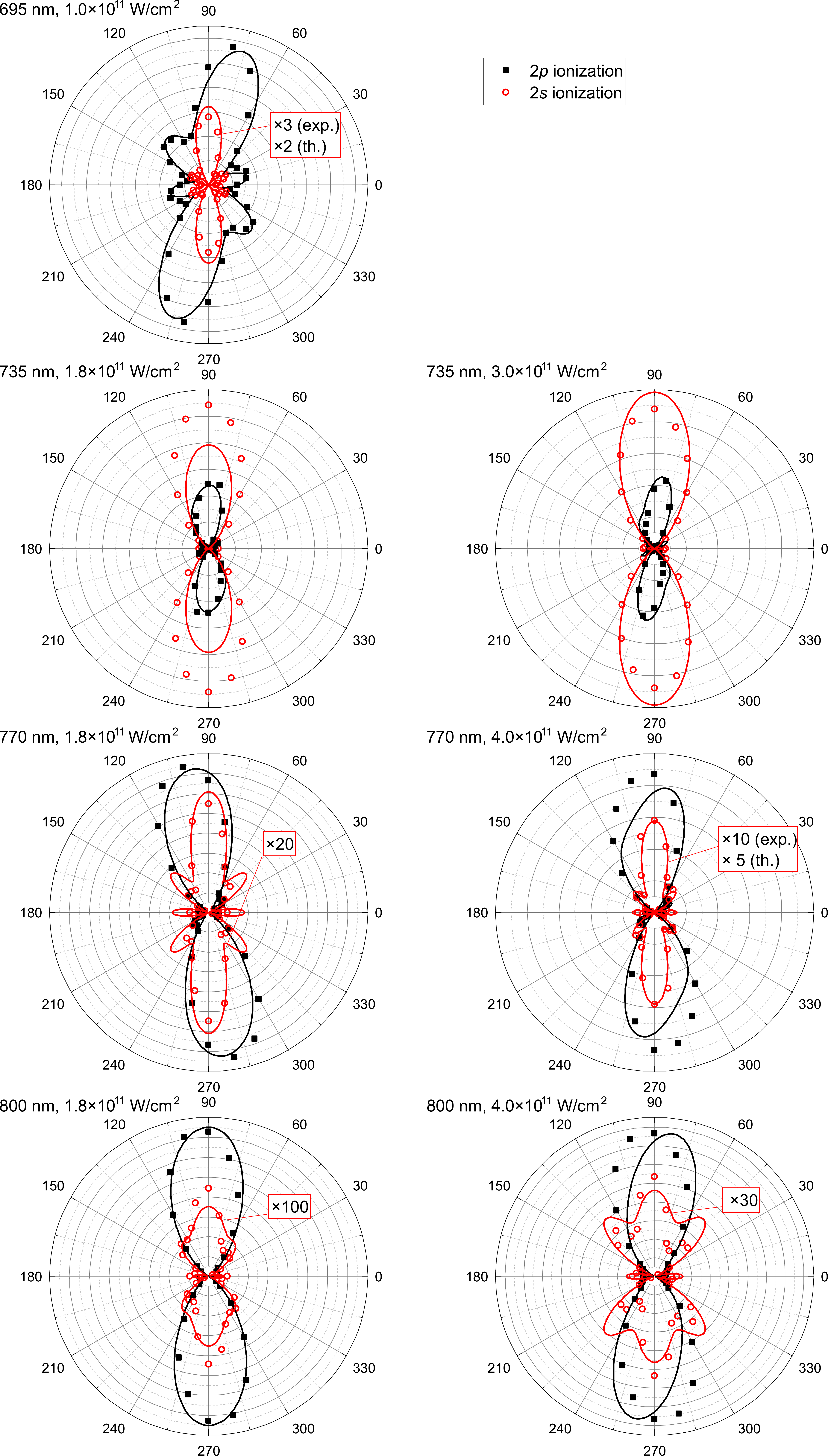}
\caption{Same as Fig.~\ref{fig:770nm180mW} (right), but for different laser wavelengths and intensities, which are 
labeled for each graph individually. Black solid squares and black lines correspond to experimental and theoretical 
results for the initial 2$p$ state, respectively. Red open circles and red lines represent the according data for 
the initial 2$s$ state. The data for the two initial states are cross-normalized in each graph, 
and -- where indicated -- multiplied by the indicated factor for better visibility.\label{fig:angdist}}
\end{figure}

According to the perturbative picture discussed above, the angular shifts observed in 
the data are sensitive to both, the relative magnitude and phase of the partial-wave amplitudes, which can change with the laser wavelength and intensity. 
 In order to  get a more complete picture of these dependences, we studied the angular distributions 
for a range of laser parameters. In Fig.~\ref{fig:angdist}, we show the cross-normalized (i.e., normalized relative to each other) spectra for 
the ionization of the 2$s$ and 2$p$ states, which have been multiplied by a suitable factor for a better visibility
when indicated.

The shape of the angular distributions agrees overall very well between measured and calculated data, with 
some moderate discrepancies at 770\,nm and 800\,nm. The relative magnitudes of the 2$s$ and 2$p$ ionization 
cross sections vary vastly over the investigated wavelength and intensity regime, and some discrepancies 
are observed as well.  They are largest for 770\,nm at 4$\times10^{11}$W/cm$^2$ (by a factor of about 2). 
As mentioned above, convolving our theoretical spectra with the experimental intensity distribution 
would likely improve the agreement, but they would also distort the clear visibility of the intensity-dependent changes.
Moreover, we have shown earlier \cite{Silva2021, Silva2021c} that our theoretical model describes the 
target system very accurately and the numerical uncertainty is extremely small. Remaining differences 
could still stem from experimental uncertainties in the laser parameters (e.g., pulse duration, 
spectrum, and intensity), which are very challenging to characterize accurately.  Here, our primary aim is not the rigorous test of our theoretical model, but rather a better understanding 
of few-photon ionization dynamics and the mechanisms at play.

All the angular distributions shown in Fig.~\ref{fig:angdist} feature two diametrically-opposed main peaks, 
aligned with the laser polarization axis for the ionization of the 2$s$ ground state and shifted 
from this axis for the ionization for the polarized (2$p$, $m=+1$) state. These angular shifts are clockwise 
for the wavelengths of 695\,nm and 735\,nm. For 770\,nm, the shifts are counterclockwise in the 
experimental spectra. In the calculation, on the other hand, the direction of the shifts flips with 
the intensity. For 800\,nm, the peaks align closely with the laser polarization axis, while the 
calculation shows a small clockwise shift for the higher intensity.  

As discussed above, the angular shifts depend sensitively on the relative magnitudes of the final 
state partial wave amplitudes. Atomic resonances can affect these magnitudes significantly. 
The most notable 1-photon resonance close to the investigated wavelength range is the $2p-3s$ resonance 
at a wavelength of 812\,nm. Because the 3$s$ state is spherically symmetric, all flux proceeding through 
this resonance will loose any information on the initial polarization direction, thereby suppressing the 
polarization of the final state. Therefore, this resonance can be expected to reduce the angular asymmetries. 
Indeed, the angular shift for a laser wavelength of 800\,nm and an intensity of 1.8$\times 10^{11}$\,W/cm$^2$ (cf.\ Fig.~\ref{fig:angdist}, bottom left) is barely noticeable. 
There are many 2-photon resonances between the 2$p$ state and higher-lying states, e.g., with $n=6$, 7, 8, and 9 at 
wavelengths of about 780\,nm, 760\,nm, 744\,nm, and 735\,nm, respectively. Here, only $p$ and $f$ states 
couple to the initial 2$p$ state due to dipole selection rules. It is difficult to pin down the effects of 
these resonances for specific laser parameters. Generally, if a $p$~state is transiently 
populated after the absorption of two photons, the set of allowed $m$ quantum numbers in the final 
state reduces to $-2$, 0, and 2, and the contribution of $m=4$ is suppressed. As a result, 
the contribution of the interference term in Eq.~(\ref{eq:dcs}) for $m=-2$ and $m'=4$, which is responsible for the six-peak structure, becomes negligible.
In contrast, a resonance to an $f$ state will allow all even $m$ quantum numbers between $-2$ and $+4$ in the final state.

\begin{figure}
\centering
\includegraphics[width=1\linewidth]{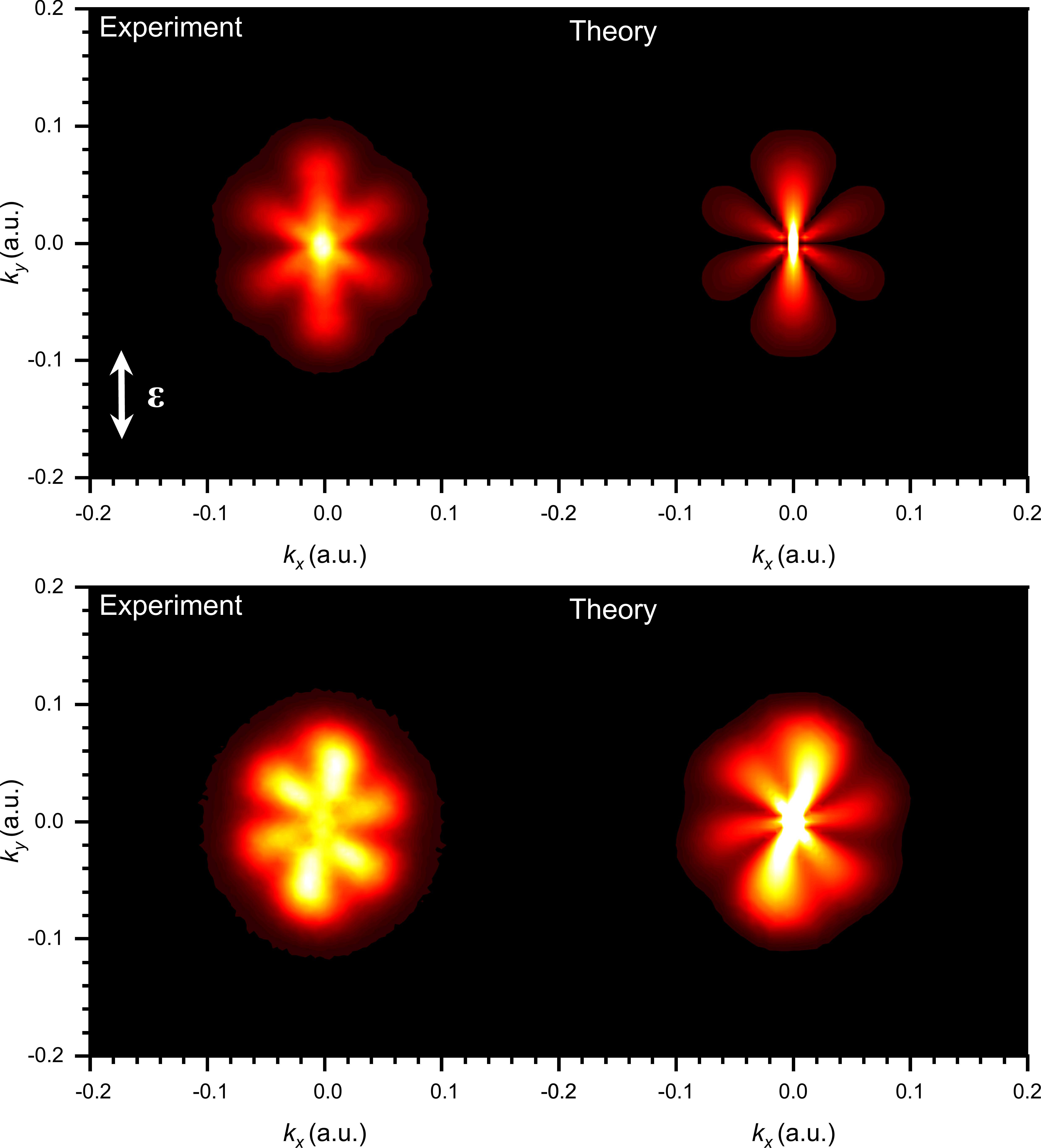}
\caption{Same as Fig.~\ref{fig:770nm180mW} (left and center), but for a laser wavelength of 695\,nm at a 
peak intensity of 1$\times 10^{11}$\,W/cm$^2$. \label{fig:695nm}}
\end{figure}

The shortest wavelength, 695\,nm, stands out in several respects: First and foremost, 
the absorption of only two photons suffices to promote the 2$p$ electron to the continuum at this wavelength. 
The ejected-electron energy is just above threshold, and the main signal from both 2$s$ and 2$p$ ionization 
is thus at very small momenta, well below 0.1\,a.u.\ (see Fig.~\ref{fig:695nm}). Furthermore, there is no 
significant resonance enhancement at this wavelength, which makes this system a particularly clean manifestation 
of the observed dichroic asymmetries. Indeed, the observed angular shift of the two dominant peaks is  
about 15$^\circ$ stronger than in all other cases investigated. The calculations reproduce the momentum 
distributions observed experimentally to excellent accuracy. 

The smaller number of absorbed photons gives rise, in LOPT, to a superposition of $p$ and $f$ waves with magnetic quantum numbers $m=-1$, 1, and 3.
As a result, the last term in the angular differential cross section of Eq.~(\ref{eq:dcs}) should vanish,
yielding only four peaks in the $\varphi$-distribution. However, this is in clear contradiction with 
our measured and calculated spectra where six peaks can be identified. While this evident 
violation of LOPT at the present comparably low intensities might be surprising, 
it can be explained by the near-resonant laser wavelength 
to the 2$s$--2$p$ transition at 671\,nm. The strong coupling between these electronic states, 
combined with the relatively large pulse length,
leads to a breakdown of LOPT even at low intensity. As a large fraction of the probability  flux passes through the 
atomic ground state, 4-photon pathways can compete with 2-photon pathways, leading 
to a significant contribution from the $(\ell=3, m=-3)$ partial wave.
The interference between the partial waves with $m=-3$ and $m=+3$ results in a 
term oscillating as $6\varphi$, i.e., the observed six-peak structure in Fig. \ref{fig:695nm}.  

\section{Summary and Conclusion}
We investigated magnetic dichroism in differential cross sections for atomic few-photon ionization 
of polarized atoms by linearly polarized femtosecond optical laser pulses. Here, dichroic asymmetries 
manifest themselves in the photoelectron angular distributions as a removal of reflection symmetry with 
respect to the laser polarization axis, and an angular shift of the main electron emission directions 
from the electric field orientation is observed. Similar asymmetries have been reported 
earlier for rather different reactions, e.g., for electron \cite{Dorn1998} or 
ion \cite{Hubele2013,GhanbariAdivi2016,GhanbariAdivi2017} impact ionization of polarized atoms.
Also for strong-field ionization, an influence of the active electron's initial angular momentum orientation on its final momentum distribution was reported \cite{Liu2018,Eckart2018}.
However, compared to these earlier studies, the present system is particularly fundamental, 
because of the well-defined energy and limited angular-momentum transfer in the multiphoton absorption process. 
We studied the dependence of the angular shift on laser wavelength and intensity, and we obtained very good 
agreement between our experimental data and calculations based on the numerical solution t
of the time-dependent Schr{\"o}dinger equation.

The observed asymmetries are qualitatively discussed in a simplified picture based on the 
electric dipole approximation in lowest order perturbation theory. Here, the final state 
is expressed as a superposition of partial waves with different orbital angular momenta $\ell$ and 
orientations $m$. Depending on the number of photons absorbed, the quantum numbers $\ell$ and $m$ 
are either all even or all odd. The dependence of the photoelectron angular distribution on 
the azimuthal angle $\varphi$ is a result of interfering partial waves with different $m$. 
The symmetry of the angular distribution with respect to the laser polarization direction is generally lifted  
if the final state features a non-vanishing mean projection of the angular momentum, i.e., $\left< m\right>\neq 0$.


In the presently studied system, the final 
polarization of the electron angular momentum  is essentially a ``remnant'' of the initial 
target orientation, which is (partially) preserved through the ionization process. 
Furthermore, several  phase-shifted partial waves contribute to the final state and interfere, resulting in the observed angular shifts.  We note that the qualitative 
explanation given here is consistent with previous analyses of elliptic dichroism in multiphoton ionization of 
unpolarized atoms \cite{Lambropoulos1988,Muller1988}, where the mean polarization $\left< m\right>$ of the final 
electron state stems from an asymmetric transfer of angular momentum by the elliptically polarized photon field. 


The general methods presented here might help to answer related questions about light-matter interaction that 
are presently under investigation. In attoclock experiments, e.g., angular asymmetries are observed in the 
tunnel-ionization regime in elliptically polarized few-cycle pulses and 
interpreted in terms of a finite time-delay of the tunneling process \cite{Eckle2008, Pfeiffer2011}. 
Future experiments involving oriented targets at much smaller Keldysh parameters than in the present study could 
be performed for linearly and elliptically polarized radiation. For both of these measurements angular shifts are expected, 
but only the latter represents an angular streaking scheme sensitive to ionization time delays. The comparison of these two situations might
shed light on open questions about the role of tunneling time delays and of phase shifts due to the target potential, 
thereby improving our understanding of the fundamentally important quantum mechanical tunneling process.

\medskip

\section*{Acknowledgments}
The experimental material presented here is based upon work supported by the National Science Foundation
under Grant \hbox{No.~PHY-1554776}.
The theoretical part of this work was funded by the NSF under
grants \hbox{No.~PHY-2012078} (M.D.\ and N.D.), \hbox{PHY-1803844} and \hbox{PHY-2110023} (K.B.), 
and by the XSEDE supercomputer allocation No.~PHY-090031.


%

\end{document}